\documentclass[%
 reprint,
 superscriptaddress,preprintnumbers,
 nofootinbib,
 amssymb,
 aps,
]{revtex4-1}

\usepackage[utf8]{inputenc}
\usepackage{hyperref}
\usepackage{amsmath,amssymb,mathtools}
\usepackage{dsfont}
\usepackage{graphicx}
\usepackage[table]{xcolor}
\usepackage{colortbl}
\usepackage{url}
\usepackage[normalem]{ulem}
\usepackage{slashed}
\usepackage[capitalise, english]{cleveref}
\usepackage[noindentafter]{titlesec} 
\allowdisplaybreaks

\usepackage{siunitx}
\usepackage{xspace}

\usepackage{multirow}

    \makeatletter
    \def\CT@@do@color{%
      \global\let\CT@do@color\relax
            \@tempdima\wd\z@
            \advance\@tempdima\@tempdimb
            \advance\@tempdima\@tempdimc
    \advance\@tempdimb\tabcolsep
    \advance\@tempdimc\tabcolsep
    \advance\@tempdima2\tabcolsep
            \kern-\@tempdimb
            \leaders\vrule
                    \hskip\@tempdima\@plus  1fill
            \kern-\@tempdimc
            \hskip-\wd\z@ \@plus -1fill }
    \makeatother

\usepackage{accents}
\DeclareMathSymbol{\widetildesym}{\mathord}{largesymbols}{"65}

\def\thesubsection{\arabic{section}.\arabic{subsection}}

\def\thesection{\arabic{section}}
\titleformat*{\subsubsection}{\normalfont \small \bfseries \boldmath}
\renewcommand{\paragraph}[1]{\vspace{.3em} \indent {\bfseries \boldmath #1 ---}\xspace }
\makeatletter
    \renewcommand{\p@subsection}{}
    \renewcommand{\p@subsubsection}{}
\makeatother

\definecolor{red}{rgb}{0.6,.0706,.1373}
\definecolor{blue}{rgb}{0,0.396,0.741}
\newcommand\myshade{80}
\colorlet{mylinkcolor}{violet}
\colorlet{mycitecolor}{violet}
\colorlet{myurlcolor}{violet}

\hypersetup{
  linkcolor  = mylinkcolor!\myshade!black,
  citecolor  = mycitecolor!\myshade!black,
  urlcolor   = myurlcolor!\myshade!black,
  colorlinks = true
}

\begin{document}


\title{The Higgs Self-Coupling at FCC-ee}

\author{Victor Maura}
\email{victor.maura\_breick@kcl.ac.uk}
\affiliation{Physics Department, King’s College London, Strand, London, WC2R 2LS, United Kingdom}

\author{Ben A. Stefanek}
\email{bstefan@ific.uv.es}
\affiliation{Instituto de F\'isica Corpuscular (IFIC), Consejo Superior de Investigaciones \\Cient\'ificas (CSIC) and Universitat de Val\`{e}ncia (UV), 46980 Valencia, Spain}

\author{Tevong You}
\email{tevong.you@kcl.ac.uk}
\affiliation{Physics Department, King’s College London, Strand, London, WC2R 2LS, United Kingdom}


\preprint{KCL-PH-TH/2025-05}

\begin{abstract}

Single Higgs production at FCC-ee probes the Higgs self-coupling at next-to-leading order (NLO). Extracting a bound requires a global analysis accounting for other possible new physics contributions up to NLO. We determine the FCC-ee sensitivity to Higgs self-coupling modifications $\delta\kappa_\lambda$ within the Standard Model Effective Field Theory (SMEFT) framework, including for the first time flavour, LEP, LHC, and FCC-ee observables in a global analysis with all leading NLO effects via one-loop renormalisation group evolution, as well as incorporating finite NLO contributions to electroweak precision and $ZH$ observables. The global sensitivity to $\delta\kappa_\lambda$ is estimated by marginalising over the effects of all other operators, bringing flavour considerations to the fore. We find that, under reasonable assumptions, FCC-ee sensitivity to $\delta\kappa_\lambda$ can exceed that of the HL-LHC.

\end{abstract}

\maketitle

\section{Introduction} 
\label{sec:intro}

The Higgs self-coupling is one of the last remaining parameters of the Standard Model (SM) to be determined precisely. Tied to the cosmological history and fate of our universe, it is of fundamental importance for a deeper understanding of the Higgs sector. While Higgs couplings measurements have reached the $\mathcal{O}(10)\%$ level~\cite{ParticleDataGroup:2024cfk, ATLAS:2022vkf, CMS:2022dwd}, the sensitivity to the Higgs self-coupling from di-Higgs production at the LHC currently stands at $\mathcal{O}(100)\%$~\cite{ATLAS:2024ish, CMS:2024ymd}. Projections for HL-LHC have been refined to an estimated $30\%$~\cite{Cepeda:2019klc,ATL-PHYS-PUB-2025-006,ATL-PHYS-PUB-2025-018}, from a combination of channels and improved analyses, using $3000$ fb$^{-1}$ of data. A percent-level determination is anticipated at future high-energy hadron colliders such as FCC-hh~\cite{gallo_higgs_2024}. 

On the other hand, future circular $e^+ e^-$ colliders at the precision frontier, such as FCC-ee~\cite{FCC:2018evy, Bernardi:2022hny, FCC-PED-FSR-Vol1} or CEPC~\cite{CEPCStudyGroup:2023quu}, would operate below the energy threshold necessary to have significant leading order (LO) access to the Higgs self-coupling in di-Higgs production. However, as pointed out in Ref.~\cite{McCullough:2013rea}, it can still be indirectly constrained through its next-to-leading order (NLO) contribution to $ZH$ single-Higgs production~\footnote{The indirect sensitivity to the Higgs self-coupling has also been studied for the single-operator case at the LHC~\cite{Gorbahn:2016uoy, Degrassi:2016wml, Bizon:2016wgr, Maltoni:2017ims, Gorbahn:2019lwq, Degrassi:2019yix, Haisch:2021hvy}, at $e^+ e^-$ colliders~\cite{Maltoni:2018ttu, Rindani:2018ubx, Rao:2021eer}, and at NNLO in electroweak precision observables~\cite{kribs_electroweak_2017,degrassi_constraints_2017}.}. The extraction of a meaningful bound must then account for other possible contributions within some theoretical framework. A well-motivated possibility is the parametrisation of heavy new physics effects using the Standard Model Effective Field Theory (SMEFT)~\cite{Weinberg:1979sa, Buchmuller:1985jz, Grzadkowski:2010es}.

An estimate of the indirect sensitivity to the Higgs self-coupling at FCC-ee marginalising over the effects of other dimension-6 operators in the SMEFT was performed in Ref.~\cite{DiVita:2017vrr}. This analysis did not vary the full set of LO operators entering the inclusive $ZH$ cross section $\sigma(ZH)$, assuming those affecting electroweak (EW) precision observables could be neglected due to strong constraints from $Z$- and $W$-pole observables. However, as recently demonstrated in Ref.~\cite{Maura:2024zxz}, the sensitivity of operators entering $\sigma(ZH)$ at LO is comparable to that of their effect at NLO at the $Z/W$-pole; the variation of those operators must therefore be taken into account simultaneously for a more complete, self-consistent global analysis. 

Since the analysis of Ref.~\cite{DiVita:2017vrr}, a full NLO computation of SMEFT contributions to $\sigma(ZH)$ inclusive has become available~\cite{Asteriadis:2024xts}. This brings sensitivity to a plethora of new operators, dominantly four-fermion operators of the type $(\bar e \gamma^\mu e) (\bar f_p \gamma_{\mu} f_p)$, where $f_p$ is any SM fermion with flavour index $p$. As these operators enter at the same order as the Higgs self-coupling, a consistent global analysis must include their variation. Sensitivity to Higgs self-coupling modifications in single-Higgs production is therefore inextricably tied to flavour symmetries that control the structure of the SMEFT~\cite{Faroughy:2020ina,Greljo:2022cah,Isidori:2023pyp}.

Higgs self-coupling modifications of $\mathcal{O}(50)\%$ correspond to a na\"ive new physics scale of about 1 TeV. However, a very specific flavour structure is required for new physics to reside at nearby energy scales~\cite{Allwicher:2023shc}. 
Flavour data and the lack of new physics beyond the SM (BSM) at the LHC have given us two important clues about this structure: i) in order to evade stringent bounds on flavour-changing neutral currents, new physics should couple universally at least to the light families, and ii) new physics does not couple strongly to valence quarks at nearby energy scales. A possibility that satisfies i) is flavour-universal new physics, described by $U(3)^5$ symmetry.
However, compatibility with both i) and ii) hints at new physics coupled universally to the light families, while it may have a larger coupling to the third generation, a scenario described by $U(2)^n$ flavour symmetries.

In this letter, we perform the first global analysis going beyond leading order to determine the projected sensitivity of FCC-ee to the Higgs self-coupling. We vary all operators entering inclusive single-Higgs production up to NLO accuracy, as well as any new operators introduced by the constraining datasets, which include flavour, EW, and collider observables. As these observables are measured at different energy scales, we consistently account for renormalisation group evolution (RGE) to build our global likelihood at a common high scale $\Lambda = 1$ TeV. We assume only the validity of the SMEFT framework and four possibilities for flavour symmetries, namely $U(3)^5$ and various $U(2)^n$ scenarios.

\section{Details of the global analysis} 
\label{sec:SMEFT}
We define the SMEFT Lagrangian as
\begin{equation}
\mathcal{L}_{\rm SMEFT} = \sum_{i} C_i(\mu) Q_i \,,
\end{equation}
where $C_i(\mu)$ are dimensionful Wilson coefficients evaluated at the renormalisation scale $\mu$ and $Q_i$ are operators in the Warsaw basis~\cite{Grzadkowski:2010es}. Throughout this work, all Wilson coefficients will be evaluated at the high scale $C_i(\Lambda)$, with $\Lambda = 1$ TeV. Due to stringent constraints on CP violation, we assume that only CP-even operators are present at the high scale $\Lambda$.

Our global analysis incorporates the observables listed in~\cref{tab:Analysis Observables}, with experimental and theoretical uncertainties implemented as in Refs.~\cite{Maura:2024zxz,Ellis:2020unq}.
All observables are computed at leading order in the SMEFT~\cite{Ellis:2020unq,Greljo:2024ytg,Stangl:2020lbh,Allwicher:2023aql,Celada:2024mcf,Allwicher:2022mcg} except for $\sigma(ZH)$ inclusive and electroweak precision observables (EWPO), which we include at NLO accuracy~\cite{Asteriadis:2024xts,Bellafronte:2023amz,Biekotter:2025nln}. Nevertheless, the leading-logarithmic NLO contributions are automatically included for all observables via the RG evolution. We solve the 1-loop RG equations using the evolution matrix approximation implemented in \texttt{DsixTools 2.0}~\cite{Fuentes-Martin:2020zaz}, which resums higher loop contributions in the leading-logarithmic series while allowing us to keep the likelihood analytic in the Wilson coefficients. We will work to $O(1/\Lambda^2)$ in the SMEFT, keeping only linear contributions to our observables. 

\begin{table}[t]

    \centering

\end{table}

\begin{table}[ht]
\begin{center}
{\def\arraystretch{1.15}
\resizebox{\columnwidth}{!}{
\begin{tabular}{|c|c|c|c|}\hline
        & Name & Description & Refs. \\\hline\hline
        \parbox[t]{3mm}{\multirow{4}{*}{\rotatebox[origin=c]{90}{{FCCee}}}}& $Z/W$-pole & Electroweak Precision Observables & \cite{Maura:2024zxz,DeBlas:2019qco}\\
        & Single $H$ & Inclusive $e^+ e^- \rightarrow ZH,\, \nu\bar{\nu} H$ cross sections & \cite{DeBlas:2019qco,Celada:2024mcf}\\
        &Diboson & Total cross sections at $163, 240, 365~{\rm GeV}$ & \cite{DeBlas:2019qco}\\
        & Di-fermion & Cross sections and $A_{\rm FB}$ at $163, 240, 365~{\rm GeV}$ & \cite{Greljo:2024ytg,deBlas:2022ofj}\\\hline
        \parbox[t]{8mm}{\multirow{2}{*}{LEP}}& Diboson & Diboson total and differential cross sections  & \cite{Ellis:2020unq}\\
        & Di-lepton & Di-lepton production for $\sqrt{s}>m_Z$ & \cite{ALEPH:2013dgf} \\\hline
        \parbox[t]{4mm}{\multirow{4}{*}{\rotatebox[origin=c]{90}{{HL-LHC}}}}&Top & $t$, $t\bar{t}$, $t\bar{t}V$, $t\bar{t}t\bar{t}$ and $b\bar{b}t\bar{t}$ (diff.) cross section& \cite{Ellis:2020unq,Ethier:2021bye}\\
        &Higgs& Higgs signal strengths and STXS data& \cite{CERNCouplingProjections2018,Stangl:2020lbh}, \cite{Ellis:2020unq}\\
        &Diboson & Fiducial differential dist. for VV and Zjj& \cite{Ellis:2020unq}\\
        &Drell-Yan & Di- and mono-lepton high-$p_{\text{T}}$ tails & \cite{Allwicher:2022gkm,Allwicher:2022mcg}\\\hline
         & Flavour & $\Delta F = 2$, $b\rightarrow c\tau\nu$, $b\rightarrow s\ell\ell$, and $b\rightarrow s\nu\nu$& \cite{Allwicher:2023shc} \\\hline\hline
        &Di-Higgs & Combined Di-Higgs signal strength& \cite{Cepeda:2019klc,ATL-PHYS-PUB-2025-006}\\\hline
    \end{tabular}
}}
\end{center}
\vspace{-0.3cm}
\caption{Datasets used in our global analysis. For all FCC-ee projections, we exclude di-Higgs data and use the latest luminosity figures in Ref.~\cite{FCC-PED-FSR-Vol1}.}
\label{tab:Analysis Observables}
\end{table}

The inclusive $ZH$ cross-section at NLO as computed in Ref.~\cite{Asteriadis:2024xts} depends on the bosonic operators $Q_H\,, Q_{H\Box}\,, Q_{HD}\,, Q_{HB/W}\,, Q_{HWB}\,, Q_W$, EW vertex corrections $\psi^2 H^2 D$, four-fermion operators of the type $(\bar e \gamma^\mu e) (\bar f_p \gamma_{\mu} f_p)$, and top Yukawa and dipole corrections~\footnote{These SMEFT contributions are enhanced by a QED K-factor of 1.19 that we include~\cite{Asteriadis:2024xts}.}. Without making any flavour assumptions there are 66 CP-even operators, the bulk of which (35 of 66) are four-fermion operators, bringing sensitivity to the flavour structure of the SMEFT. Our approach is to analyse sensitivity to Higgs self-coupling modifications assuming four flavour symmetries that allow for new physics at nearby energy scales, shown in~\cref{tab:flavSyms}. Motivated by tight LHC bounds on new physics coupling to valence quarks, the scenario denoted ``$U(2)^5$ (3rd-gen. dominance)" corresponds to the $U(2)^5$ scenario with the additional dynamical assumption that couplings to the light generations are suppressed. This is implemented by keeping only $U(2)^5$-invariants with fully third-generation fields. 
\begin{table}[ht]
\begin{center}
{\def\arraystretch{1.1}
\begin{tabular}{|c|c|}\hline
        Flavour symmetry & CP-even parameters  \\\hline\hline
        $U(3)^5$ & 41  \\\hline
        $U(2)_q \times U(2)_u \times U(3)^3$ & 72 \\\hline
        $U(2)^5$ & 124 \\\hline
        $U(2)^3_{q,u,d} \times U(1)^3_{e,\mu,\tau}$ & 168 \\\hline
        $U(2)^5$ (3rd-gen. dominance) & 53  \\\hline
    \end{tabular}
}
\end{center}
\caption{Flavour symmetries considered in this work. We have defined $U(3)^3 \equiv U(3)_d \times U(3)_l \times U(3)_e$.}
\label{tab:flavSyms}
\end{table}

All symmetries including flavour and CP are imposed only at the high scale $\Lambda$---we consistently allow them to be broken by rotations to the SM fermion mass basis as well as RG effects. Even our $U(3)^5$ scenario thus includes Minimal Flavour Violating effects~\cite{DAmbrosio:2002vsn} at 1-loop. The rotation matrices are fixed assuming the most conservative ``down-aligned" scenario discussed in Ref.~\cite{Allwicher:2023shc}, where flavour-violating effects in the more sensitive down-quark observables come only from RG contributions.

The four-fermion operators entering $\sigma(ZH)$ at NLO may be constrained by $e^+ e^- \rightarrow \bar f f$ observables above the $Z$-pole at LEP and FCC-ee, such as total cross sections and forward-backward asymmetries $A_{\rm FB}$~\footnote{We use the $A_{\rm FB}$ projections for FCC-ee from~\cite{deBlas:2022ofj}, which do not include strange or top quarks. For those we perform our own projection using the optimal tagging working point of~\cite{Greljo:2024ytg}.}. In the case of FCC-ee, the 365 GeV run enables constraints on $eett$ operators. Semileptonic operators receive complementary constraints from high-energy Drell-Yan tails at the LHC. To constrain some of the bosonic operators, we include data from single Higgs production and decay at the LHC, as well as diboson data from both LHC and FCC-ee. 
Wherever LHC data is concerned, we take projections for the HL-LHC from official publications if available~\cite{CERNCouplingProjections2018,cepeda_higgs_2019}, otherwise we rescale the Run 2 statistical uncertainties by the luminosity increase to $3~{\rm ab}^{-1}$ and assume a factor of two reduction in systematic and theoretical uncertainties, following the \textit{S2}-scenario described in Ref.~\cite{Cepeda:2019klc}.
While subleading, we also include the  $e^+ e^- \rightarrow \bar\nu\nu h$ single Higgs production channel at FCC-ee. Concerning FCC-ee projections, we use the luminosity figures presented in the Feasibility Study Report~\cite{FCC-PED-FSR-Vol1}, which are roughly double for all runs above the $Z$-pole compared with Ref.~\cite{Bernardi:2022hny}.  We keep and vary all additional operators entering the aforementioned observables at LO in our global analysis. 

As previously mentioned, variations in EW vertices and the bosonic operators $Q_{HD}$ and $Q_{HWB}$ must be taken into account, as EWPO are sensitive to many $ZH$ operators at NLO with a similar precision as $\sigma(ZH)$ at leading order. Indeed, EWPO at NLO are sensitive to almost all parameters in our fit. 

Therefore, while we always include the full LO dependence, we make the following assumptions \emph{which apply only concerning EWPO at NLO}:
\begin{enumerate}
\item We keep the full NLO dependence for operators we already had in the observables discussed above.
\item New operator dependencies at NLO proportional to weak couplings are neglected. We include new operators only if they are proportional to the large couplings $y_t$ or $g_s$.
\end{enumerate}

This assumption leads to the inclusion of $4t$, $2t2q$ and $2t2l$ operators while it only excludes third-family $4l$ and light-quark $4q$ operators that do not enter any other observables added so far. To constrain the new four-quark and semi-leptonic operators involving $q_L^3$ and $t_R$, we add top data projections for the HL-LHC~\cite{Ellis:2020unq,Ethier:2021bye} and low-energy flavour data from $B,K,D$ meson mixing as well as semileptonic $b\rightarrow c\tau\nu$, $b\rightarrow s\ell\ell$, and $b\rightarrow s\nu\nu$ transitions. We include all flavour observables of these types given in Ref.~\cite{Allwicher:2023shc}, which does not add any significant new operator dependence. On the other hand, we consistently include all new operators brought by the top dataset (including their full NLO contributions to EWPO). 

In total, 150 operators are included in our analysis before imposing any flavour symmetry, which is roughly half of all flavour-conserving and CP-even operators in the SMEFT. In the limit of $U(2)^3_{q,u,d} \times U(1)_{e,\mu,\tau}$-invariance, 116 independent Wilson coefficients remain, which is the largest parameter set we consider. Even in this case, the more than 400 observables we include is sufficient to close our global fit, though some linear combinations of Wilson coefficients that do not enter single-Higgs production remain poorly constrained in the least symmetric cases. To ensure the validity of our dimension-6 EFT, as well as to justify neglecting quadratic contributions to observables, we place an upper bound on the size of all Wilson coefficients except $C_H$ in the fit. This is implemented by the inclusion of a ``boundary condition (BC) likelihood" 
\begin{equation}
\chi^2_{\rm BC} = 4\sum_{i\neq H} \left(\frac{C_i}{C_{\rm BC}} \right)^2\,, \hspace{5mm} C_{\rm BC} = \frac{1}{(250~{\rm GeV})^2}\,,
\end{equation}
where the sum runs over all Wilson coefficients in the fit except for $C_H$. This likelihood has the effect of enforcing $C_i < C_{\rm BC}$ at 95\% CL, essentially acting as a theoretical prior on the size of the Wilson coefficients. We do not impose a boundary condition for $C_H$ as our goal is to determine the FCC-ee sensitivity to this Wilson coefficient. Instead, $C_H < C_{\rm BC}$ as determined from our global fit will serve as a consistency check on the analysis. 

The choice of $C_{\rm BC}$ is motivated by considering new physics (NP) with a mass scale of $M_{\rm NP} \gtrsim 1$ TeV to trust the EFT description, while assuming a perturbative NP coupling $g_{\rm NP} < \sqrt{4\pi}$. Together, these conditions yield $C_{\rm BC} = g_{\rm NP}^2/M_{\rm NP}^2 \lesssim 1/(250~{\rm GeV})^2$, corresponding to new physics effects smaller than SM electroweak processes.

We build our likelihood by performing RGE to express all observables $O_i$ as linear, analytic functions of the high-scale Wilson coefficents $C_i(\Lambda)$, which we use to compute the $\chi^2$-function
\begin{equation}
\chi^2_{\rm data} = \sum_{ij}[O_{i,\text{exp}}-O_{i,\text{th}}] (\sigma_{\rm exp}^{-2})_{ij}[O_{j,\text{exp}}-O_{j,\text{th}}] \,.
\end{equation}
We then add the boundary condition likelihood such that the total likelihood is given by $\chi^2 = \chi^2_{\rm data}+ \chi^2_{\rm BC}$. 

\section{Higgs self-coupling sensitivity at FCC-ee} 

\begin{figure*}[t]
  \centering
\includegraphics[width=0.85\textwidth]{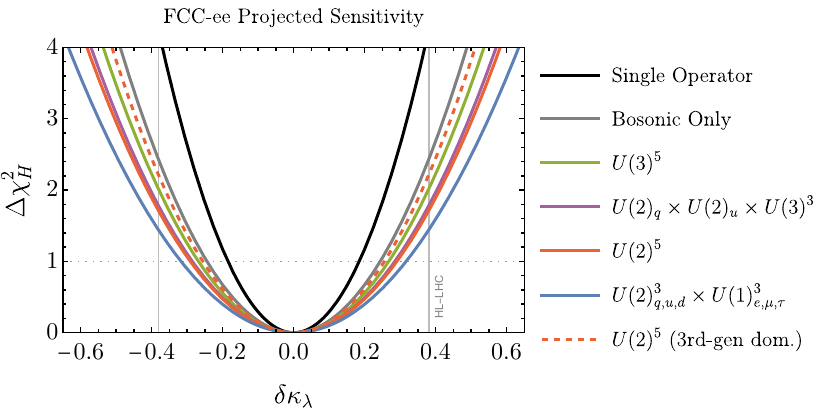}
  \caption{Marginalised SMEFT likelihood for the Higgs self-coupling, $\delta\kappa_\lambda$, under various assumptions for flavour symmetries. The single-operator and bosonic-only scenarios are also shown for reference. For comparison, vertical solid grey gridlines denote the fully marginalised HL-LHC sensitivity at 68\% CL.}
  \label{fig:chi2plot}
\end{figure*}

We are interested in modifications of the Higgs self-coupling, which we parametrise by $\kappa_\lambda \equiv \lambda_3 / \lambda_{3}^{\rm SM}$, where 2$\lambda_{3}^{\rm SM} v^2 = m_h^2 $. The shift $\delta\kappa_\lambda$ at the dimension-6 level from the operator $(H^\dagger H)^3$ with Wilson coefficient $C_H$ is
\begin{equation}
\delta \kappa_\lambda \equiv \frac{\lambda_3}{\lambda_{3}^{\rm SM}} - 1 = \frac{v^2 C_H}{\lambda_{3}^{\rm SM}} \,.
\label{eq:DKL}
\end{equation}
We will always determine $\delta \kappa_\lambda$ using~\cref{eq:DKL} with $C_H$ evaluated at $\Lambda = 1$ TeV.
Since our likelihood is Gaussian and fully analytic in the Wilson coefficients, we extract the allowed range for $C_H$ marginalised over the contributions of all other Wilson coefficients by analytically computing and inverting the Hessian matrix ${\bf H}$
\begin{equation}
{\bf H}_{ij} \equiv \frac{1}{2}\frac{\partial^2 \chi^2}{\partial C_i \partial C_j}\, , \hspace{5mm} \sigma_i^2 = {\rm diag}({\bf H}^{-1})\,,
\end{equation}
where $\sigma_i$ is the marginalised 68\% CL interval for each Wilson coefficient $C_i$. The results of our analysis are shown in~\cref{tab:CHvals}, where we report $\sigma_H$ as well as the corresponding 68\% CL interval on $\delta\kappa_\lambda$, for several different flavour assumptions. 
\begin{table}[ht!]
\begin{center}
{\def\arraystretch{1.2}
\resizebox{\columnwidth}{!}{
\begin{tabular}{|c|c|c|}\hline
        Scenario & $\sigma_H [{\rm TeV}^{-2}]$ & 68\% CL $\delta\kappa_\lambda$ \\\hline\hline
        $C_H$ Only & 0.39 & 18\% \\\hline
        Bosonic Only & 0.52 & 24\% \\\hline
        $U(3)^5$ & 0.57 & 27\% \\\hline
        $U(2)_q \times U(2)_u \times U(3)^3$ & 0.61 & 29\% \\\hline
        $U(2)^5$ & 0.62 & 29\% \\\hline
        $U(2)^3_{q,u,d} \times U(1)^3_{e,\mu,\tau}$ & 0.68 & 32\% \\\hline
        $U(2)^5$ (3rd-gen. dominance) & 0.54 & 25\% \\\hline
    \end{tabular}
}}
\end{center}
\caption{Higgs self-coupling sensitivities at FCC-ee switching on $C_H$ only, bosonic operators only, or imposing the flavour symmetries listed in~\cref{tab:flavSyms}.}
\label{tab:CHvals}
\end{table}
We also plot the marginalised likelihood for $C_H$ (or $\delta\kappa_{\lambda}$) under the various flavour scenarios in~\cref{fig:chi2plot}. For comparison, the vertical gridlines give the 68\% CL marginalised sensitivity at the HL-LHC, which we find to be $\delta\kappa_\lambda = 0.38$ from a global fit combining di-Higgs production with all other HL-LHC observables in~\cref{tab:Analysis Observables}~\footnote{The marginalised HL-LHC sensitivity does not depend on the choice of flavour symmetry as long as it is relaxed enough to allow variations in all operators entering double-Higgs production. This is true for all symmetries we consider except $U(3)^5$. }. For the projected di-Higgs sensitivity at the HL-LHC, we use the $S2$ scenario of Ref.~\cite{ATL-PHYS-PUB-2025-018} with 3 $\text{ab}^{-1}$, which yields $\delta\kappa_\lambda = 0.36$ for a $C_H$-only fit. The apparent discrepancy with the quoted $\delta\kappa_\lambda = 0.3$ is due to the running of $C_H$ up to $\Lambda = 1$ TeV.  We emphasize that di-Higgs data is not included in our global fit for FCC-ee. Our key result, shown in~\cref{fig:chi2plot}, is that FCC-ee is able to achieve better sensitivity on $\delta\kappa_\lambda$ than the HL-LHC for all flavour symmetries considered in our analysis, namely $\delta\kappa_\lambda \lesssim 30\%$.

To examine the dependence of our results on the choice of 
the theoretical prior imposing an upper limit on all Wilson coefficients, we vary $C_{\rm BC}$ in the range $(350~{\rm GeV})^{-2} < C_{\rm BC} <  (150~{\rm GeV})^{-2}$. Under this variation, $\sigma_H$ changes by about $5\%$ in the $U(2)_{q,u,d} \times U(1)^3_{e,\mu,\tau}$ scenario, while it is $\lesssim 2\%$ in all others. If the boundary condition is removed entirely, we find $\delta\kappa_\lambda \lesssim 37\%$ at 68\% CL in the least symmetric case, demonstrating the stability of our fit.

Restoring the boundary condition and examining the two least symmetric cases $U(2)^5$ and $U(2)_{q,u,d}^3 \times U(1)^3_{e,\mu,\tau}$, we find the most poorly constrained directions involve four-fermion operators with all 3rd-generation fields that do not enter single-Higgs production at NLO, namely $Q_{qd}^{(1)},Q_{ud}^{(1)}, Q_{qu}^{(1)}, Q_{qu}^{(8)}, Q_{ld}, Q_{ed}, Q_{eu}, Q_{lu}$, which have poorly-lifted flat directions in our datasets. Of the operators entering in single-Higgs production, the weakest constraints are on the top Yukawa and EW vertex correction $[C_{uH}]_{33},[C_{Hu}]_{33} \lesssim 1/({\rm 900~GeV})^2$ at 95\% CL. While $eett$ operators are allowed the most variation of the semileptonics, they are overall well constrained in our fit---the weakest bounds are $[C_{l,eu}]_{1133} \lesssim 1/({\rm 1.4~TeV})^2$ at 95\% CL. This is because a flat direction in $\sigma(e^+ e^- \rightarrow t \bar t)$ at 365 GeV is broken by the $A_{\rm FB}^t$ as well as the large NLO contribution of these operators to EWPO proportional to $y_t^2 N_c$.


\section{Implications for BSM models} 
\label{sec:model}
We now discuss the implications of our results in the context of explicit BSM scenarios. Interestingly, Table~\ref{tab:CHvals} shows that the sensitivity in the $U(2)^5$ scenario can become better than $U(3)^5$, $\delta\kappa_\lambda \lesssim 25\%$, if new physics is dominantly coupled to the 3rd generation. This is the case for a large swathe of well-motivated new physics scenarios addressing the flavour and/or EW hierarchy problems, such as flavour deconstructions~\cite{Fuentes-Martin:2022xnb,Davighi:2022fer,Davighi:2023iks,Davighi:2023evx,Davighi:2023xqn,Barbieri:2023qpf,Capdevila:2024gki,Fuentes-Martin:2024fpx,Lizana:2024jby}, composite Higgs, and split-supersymmetric models. For example, the class of composite Higgs models where the top quark has a significant degree of compositeness as studied in Ref.~\cite{Stefanek:2024kds} are well described by $U(2)^5$ with 3rd-generation dominance.

In general, it seems difficult to construct explicit models naturally yielding sizeable Higgs self-coupling modifications without introducing correlated effects in other better constrained operators~\cite{Durieux:2022hbu, Bhattiprolu:2024tsq}. Roughly speaking, a 1-loop hierarchy is allowed without fine-tuning, which is best exemplified by the custodial quadruplet model generating only $Q_H$ at tree level~\cite{Durieux:2022hbu}. The full 1-loop SMEFT matching was given in Ref.~\cite{Maura:2024zxz}, which generates $Q_{H\Box}$, operators corresponding to the EW $\hat{W}$ and $\hat{Y}$ parameters, and $Q_{uH},Q_{dH},Q_{eH}$ proportional to their respective SM Yukawa couplings. The model thus inherits the same flavour structure as the SM Yukawas, making it well described by all our flavor symmetries except $U(3)^5$; one may therefore expect a sensitivity of at least $\delta\kappa_\lambda \lesssim 25\%$ at FCC-ee.

Even in this model designed to only generate $Q_H$ at tree level, the loop-generated operators lead indirectly to comparable or better constraints on $\delta\kappa_\lambda$ than those coming from $Q_H$, both at HL-LHC and FCC-ee. In the HL-LHC case, there is comparable sensitivity in $H\rightarrow b\bar b$ decays modified at 1-loop by $Q_{dH}$, while in the case of FCC-ee the best constraint comes from the large RG mixing of $Q_{H\Box}$ into the custodial-violating operator $Q_{HD}$ corresponding to the $\hat T$ parameter~\cite{Maura:2024zxz}. As these operators are fully correlated with $Q_H$ within this model, measurements of these observables would constrain the model parameters to corresponding limits of $\delta\kappa_\lambda \lesssim 40\%$ (HL-LHC) and $\delta\kappa_\lambda \lesssim 14\%$ (FCC-ee), which are similar or better than their respective single-operator sensitivity on $Q_H$ in both cases. This example does not provide a no-go theorem---large Higgs self-coupling modifications with suppressed effects elsewhere may still be possible in certain BSM models.

\section{Outlook and Conclusions} 
\label{sec:conc}
The LHC has opened a new frontier to the Higgs sector and TeV scale that we have barely begun to explore. Many open questions surround the Higgs: Is it elementary or composite? What generates its potential and Yukawa couplings? What stabilises the electroweak scale against quantum corrections?  Was the electroweak phase transition first- or second-order? Are there any other particles that get most of their mass from the Higgs? Is the electroweak vacuum metastable? At what scale is it destabilised? 

All these fundamental questions and many more can be explored at a next-generation Higgs factory. In particular, the Higgs self-coupling is a major target for future colliders. At FCC-ee, the NLO contribution of the Higgs self-coupling means that the marginalised sensitivity greatly benefits from the plethora of complementary measurements across FCC-ee, together with the HL-LHC and flavour sectors. In this letter, we provided the first self-consistent assessment of the indirect sensitivity to Higgs self-coupling modifications $\delta\kappa_\lambda$ at FCC-ee via a global SMEFT analysis, including all the leading NLO effects. 

The large number of flavoured operators at NLO are controlled by imposing various flavour symmetries that may be exhibited by new physics lying within reach of future colliders. These symmetries are realized in a wide class of well-motivated BSM models, including those deeply connected to the nature of the electroweak scale such as composite Higgs or supersymmetric models. In all cases, we found that FCC-ee can achieve a global sensitivity which is better than di-Higgs production at HL-LHC. It may considerably outperform the HL-LHC in constraining explicit models and for BSM physics exhibiting a Yukawa-like flavour structure dominantly coupled to the 3rd generation, where we find a marginalised global sensitivity of $\delta\kappa_\lambda \lesssim 25\%$. FCC-ee can therefore reach an impactful level of sensitivity on the Higgs self-coupling~\cite{Bhattiprolu:2024tsq}. This conclusion does not depend strongly on the particular flavour symmetry, as overall we find $\delta\kappa_\lambda \lesssim 30\%$.

There is a clear consensus on a Higgs factory as the next step for particle physics~\cite{CERN-ESU-015, osti_2368847}. The importance of complementary and interdependent measurements across all sectors of the Standard Model not directly related to the Higgs must also not be overlooked; in particular, the crucial role of a Tera-$Z$ factory for exploring BSM physics at the tens of TeV scale has recently been emphasised e.g. in Refs.~\cite{Allwicher:2023aql,Allwicher:2023shc,Stefanek:2024kds,Allwicher:2024sso, Erdelyi:2024sls, Erdelyi:2025axy, Ge:2024pfn, Gargalionis:2024jaw,Maura:2024zxz}. As a Tera-$Z$, Higgs, and flavour factory, FCC-ee provides the most complete coverage of a broad range of particle physics phenomena, ranging from precision measurements at the $Z/W$-pole, $WW$, $ZH$, and $t\bar{t}$ thresholds, followed by direct exploration of the high-energy frontier at FCC-hh that will ultimately probe the Higgs self-coupling at the percent level.

\emph{Note added:} After our preprint was released on the arXiv, Ref.~\cite{terHoeve:2025hfx} appeared which also studies the sensitivity to the Higgs self-coupling at FCC-ee. The main differences seem to be: 1) Their analysis omits the majority of relevant semileptonic and 4-lepton operators that are included in our global fit and cannot be neglected by the assumed symmetries; 2) While we use the inclusive $ZH$ cross-section, their limits are dominated by $\sigma(ZH) \times \text{Br}(H\rightarrow \bar b b)$, which is expected to be measured with a higher precision of 0.21\% vs. 0.3\% for the inclusive cross section. We conservatively chose the latter as the NLO SMEFT contributions to the inclusive cross section have been calculated and can be included for a fully self-consistent global analysis at NLO, whereas the $\sigma(ZH) \times \text{Br}(H\rightarrow \bar b b)$ observable requires NLO SMEFT contributions to $\Gamma({H\rightarrow \bar b b})$ as well as to the total Higgs width, which are not included in Ref.~\cite{terHoeve:2025hfx}.

\section*{Acknowledgments}
We thank Gauthier Durieux, Pier Paolo Giardino, and Tamara Vázquez-Schröder for helpful feedback, and Alejo Rossia for discussions. The work of BAS is supported by a CDEIGENT grant from the Generalitat Valenciana with grant no. CIDEIG/2023/35. VM is supported by a KCL NMES faculty studentship. TY is supported by United Kingdom Science and Technologies Facilities Council (STFC) grant ST/X000753/1. BAS and VM thank CERN for their hospitality while part of this work was carried out.

\appendix 
\renewcommand{\thesection}{\Alph{section}}
\renewcommand{\thesubsection}{\Alph{section}.\arabic{subsection}}
\setcounter{section}{0}

\bibliographystyle{JHEP}
\bibliography{refs.bib}

\end{document}